\documentclass[manuscript]{aastex}

\usepackage[]{graphicx}

\shorttitle{Hectospec Velocity Dispersion}
\shortauthors{Fabricant et al.}

\begin{document}



\title{Measuring Galaxy Velocity Dispersions with Hectospec}


\author{Daniel Fabricant, Igor Chilingarian, Ho Seong Hwang, Michael J. Kurtz and Margaret J. Geller}
\affil{Center for Astrophysics, Cambridge, MA 02138}
\author{Ian P. Del'Antonio} \affil{Brown University, Providence, RI 02912}
\author{Kenneth J. Rines} \affil{Western Washington University, Bellingham, WA 98225}



\begin{abstract}
We describe a robust technique based on the ULySS IDL code for measuring
velocity dispersions of galaxies observed with the MMT's fiber-fed
spectrograph, Hectospec.  This procedure is applicable to all
Hectospec spectra having a signal-to-noise $\ga$5 and weak emission
lines.  We estimate the internal error in the Hectospec velocity
dispersion measurements by comparing duplicate measurements of 171
galaxies.  For a sample of 984 galaxies with a median z=0.10, we
compare velocity dispersions measured by Hectospec through a
1.5$^{\prime\prime}$ diameter optical fiber with those measured by the
Sloan Digital Sky Survey (SDSS) and Baryon Oscillation Spectral Survey
(BOSS) through 3$^{\prime\prime}$ and 2$^{\prime\prime}$ diameter optical
fibers, respectively.  The systematic differences between the Hectospec
and the SDSS/BOSS measurements are $<$7\% for velocity dispersions between
100 and 300 km s$^{-1}$, the differences are no larger than the differences among
the three BOSS velocity dispersion reductions.  We analyze the scatter
about the fundamental plane and find no significant redshift
dependent systematics in our velocity dispersion measurements to z$\sim$0.6.
This analysis also confirms our estimation of the measurement errors.
In one hour in good conditions, we demonstrate that we achieve 30 km s$^{-1}$
velocity dispersion errors for galaxies with an SDSS r fiber magnitude of 21.

\end{abstract}




\section{INTRODUCTION}

\label{sect:intro}  

Measurements of stellar velocity dispersions in galaxies have broad application
including estimation of galaxy masses, galaxy classification, and distance
measurements.  The use of velocity dispersion measurements to determine
galaxy properties and their evolution requires a clear understanding of the
statistical and systematic errors in these measurements
over a wide redshift range.  Comparison
of large samples of measurements obtained with different instruments and
different techniques constrains these errors.

These broad applications of stellar velocity dispersion in galaxies
motivate our investigation of velocity dispersion measurements from
moderate to low signal-to-noise (SN) spectra originally obtained
for redshift measurements.  Hectospec, the MMT's fiber-fed
spectrograph \citep{fab05}, has obtained spectra of $\sim$700,000 unique
objects; an appreciable fraction of these spectra can yield reliable
velocity dispersions.  Hectospec's fibers subtend 1.5$^{\prime\prime}$
on the sky.  Most spectra, obtained with a 270 line mm$^{-1}$
grating, have ~5 \rm\AA~FWHM resolution.  We discuss results
obtained with direct spectral fitting, an approach first described by
\citet{rix92}.  Direct spectral fitting is conceptually
straightforward and allows simple masking of portions of the spectrum
contaminated with strong skylines, bad pixels, or emission lines.  We
use a IDL-based software package, ULySS, developed by \citet{kol09} to
perform direct fitting of Hectospec spectra.  ULySS fits observed
spectra to model spectra of synthesized galaxy stellar populations
broadened by the instrumental line spread function and a velocity
dispersion.

An extensive literature reporting velocity dispersion techniques and
measurements follows the pioneering work of
\citet{min54} who reported measurements of M31's velocity dispersion.
The earliest measurements using optical fiber front-ends were
performed at the AAT \citep{col87,luc88}.  \citet{jor95} made
extensive comparisons of velocity dispersions from the fiber-fed
OCTOPUS spectrograph with slit spectrograph measurements, and obtained
consistent results.  Beginning with the first data release \citep{aba03}, the Sloan
Digital Sky Survey (SDSS) has made available extensive catalogs of well-calibrated
velocity dispersions obtained with fiber-fed spectrographs.  The only serious
disadvantage of measuring velocity dispersions through fibers rather
than directly with a slit spectrograph is the sacrifice of spatial resolution
for multiplex advantage.

In Section 2 we describe application of ULySS to Hectospec spectra
with weak nebular line emission allowing uncontaminated measurement of
stellar absorption features.  We calculate the internal errors in our
velocity dispersion measurements in Section 3.  In Section 4 we
compare our velocity dispersion measurements for 984 galaxies with
SDSS/Baryon Oscillation Spectral Survey (BoSS) Date Release 9 (DR9)
pipeline measurements.  We plot the scatter about the fundamental plane for
a sample of 1857 galaxies to a redshift of 0.6 in Section 5 to demonstrate
that our velocity dispersion errors are accurate at higher redshifts.
We describe how to plan velocity dispersion measurements with Hectospec
in Section 6, and give our conclusions in Section 7.  We adopt cosmological
parameters H$_0$=70, $\Omega_M$=0.3, and $\Omega_\Lambda$=0.7.


\section{Hectospec Velocity Dispersion Measurements}


\subsection{Introduction}

The ULySS algorithms \citep{kol09} are based on an earlier IDL
package, NBURSTS, developed by \citet{chi07}. NBURSTS in turn is based
on the pPXF package developed by \citet{cap04}.  ULySS simultaneously
fits a spectrum with the internal galaxy dynamics and parameters describing a star
formation history.  To extract accurate
velocity dispersions, the model spectrum must be convolved to the same
wavelength dependent spectral resolution as the data. ULySS then
applies a multiplicative polynomial to account for errors in flux
calibration of the data and for inaccuracies in the model prediction.

The galaxy spectral models we use are single age stellar populations (SSP)
parameterized by age and metallicity.  We used a precomputed grid of
SSP models \citep{pru11} calculated with the PegaseHR code
\citep{leb04} from the MILES stellar library \citep{san06}, assuming a
Salpeter initial mass function and solar neighborhood abundances. This
grid of models is available on the ULyss web site
(ulyss.univ-lyon1.fr).  We adopt a cosmology with $H_0$=70, $\Omega_M$=0.3,
and $\Omega_\Lambda$=0.7.

\subsection{ULySS Parameters}

The first step in our analysis is relative flux
calibration of the Hectospec spectra following the techniques
described in \citet{fab08}.  We write out the deredshifted fluxed spectra
in a FITS format compatible with ULySS.  As
described in \citet{fab08}, the flux calibration is quite stable over
time, and the ULySS multiplicative polynomial accounts for any small
errors in flux calibration.

We experimented extensively with the
available ULySS parameters.  We obtain the lowest velocity dispersion errors and
stablest results by restricting the spectral range to
4100-5500 \rm\AA, by restricting the model metallicities (log(${model}\over{solar}$))
between -0.5 and 0.5, and by using a third order multiplicative polynomial.
Accessing lower metallicities in model fits allows an unphysical
degeneracy between age and metallicity for low SN spectra \citep{wor95}.  Higher
order multiplicative polynomials do not meaningfully reduce
chi squared and sometimes attempt to null real spectral features
on low SN spectra.

\subsection{Hectospec's Line Spread Function}

We determine Hectospec's line spread function (LSF) using the ULySS
routine uly\_lsf.  This routine derives the line spread function by
fitting twilight flat spectra to a high resolution model (R=10000) of
the solar spectrum provided in the ULySS distribution.  As supplied,
uly\_lsf fits only a velocity shift and a Gaussian LSF, but is
easily modified to fit h3 and h4 Hermite polynomials to describe a
non-Gaussian LSF.  Although Hectospec's optical fibers do provide a
somewhat flat topped LSF, measurements of h3 and h4 with uly\_lsf
scatter closely about 0.  We therefore fix the h3 and h4 terms at
zero.

Hectospec's fiber feed guarantees a consistent input aperture, and the
line spread function should be quite stable; the only time
dependent changes should arise from focus variations and changes in
the fiber focal ratio degradation.  The Hectospec focus is regularly
checked and is maintained within a tight range, and Hectospec was
carefully designed to minimize focal ratio degradation \citep{fab05}.
In addition to time dependent changes, variation in the image quality
of Hectospec's optics and the flatness and alignment of the CCD
detectors can introduce fiber to fiber variations in the line spread
function.  These spatially and time dependent variations in the line
spread function can be recovered from the twilight flats for each
night of data and each fiber. A new pipeline under development
will correct for these issues.  We show that very acceptable results
can be obtained from the current pipeline by
using a line spread function averaged over fiber and time.

We have measured the Hectospec line spread function for each of the
300 fibers on three randomly chosen nights: 13 October 2007, 20
November 2008, and 15 October 2009.  For these 900 measurements we
calculate the LSF in 17 wavelength bins, each 200 \rm\AA~wide,
centered between 3800 to 8900 \rm\AA.  For each bin, we calculate the
mean Gaussian (1 $\sigma$) LSF in km s$^{-1}$ and the standard
deviation in the LSF.  The LSF ranges between 172 km s$^{-1}$ at 3800
\rm\AA, 105 km s$^{-1}$ at 6000 \rm\AA, and 78 km s$^{-1}$ at 9000
\rm\AA, or 5.1, 4.9, 5.5 \rm\AA~FWHM, respectively.  The standard
deviation in these measurements is typically 3 km s$^{-1}$, or 2\% of
the LSF at the blue end and 5\% of the LSF at the red end of
Hectospec's spectral range.  These standard deviations are small
enough to support use of the average LSF.  Spectral
regions affected by strong atmospheric absorption (particularly
between 6800 and 7600 \rm \AA) yield incorrect LSFs from the
twilight flats and must be rejected.  We fit a third order polynomial
to the valid data points; the coefficients of the fit are in
Table~\ref{lsf}.

\begin{figure}
\begin{center}
\begin{tabular}{c}
\includegraphics[height=6in]{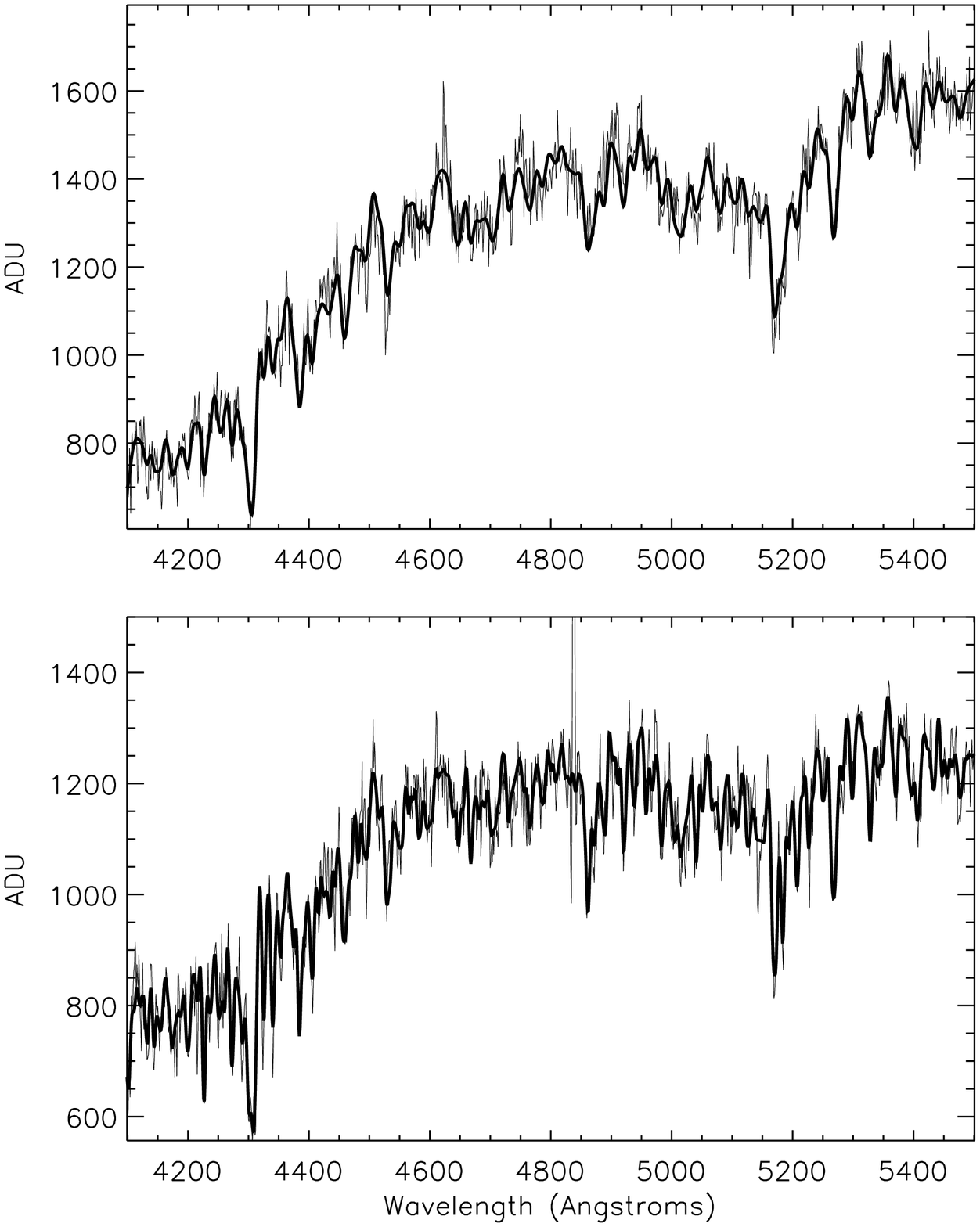}
\end{tabular}
\end{center}
\caption{\label{ulyss}  Sample ULySS fits (bold line) overplotted on Hectospec spectra (fine line). The spectrum
in the bottom panel yields a velocity dispersion of 101$\pm$6 km s$^{-1}$ and the spectrum
in the top panel yields a velocity dispersion of 281$\pm$8 km s$^{-1}$.}
\end{figure}

To analyze the fluxed Hectospec spectra
shifted to rest frame wavelengths, we use the LSF appropriate to
the original observed wavelengths. In addition, we correct for the
intrinsic resolution of the MILES stellar library, 2.51 \rm\AA~FWHM
\citep{fal11}, or $\sigma$=1.066 \rm\AA~(see also \cite{pru11,bei11}).
We subtract this resolution
in quadrature from the redshift-shifted Hectospec LSF to produce a
final LSF for each spectrum.  Figure~\ref{ulyss} shows sample Hectospec
spectra with the ULySS fits overplotted.

\section{Internal Errors}

\begin{figure}
\begin{center}
\begin{tabular}{c}
\includegraphics[height=6in]{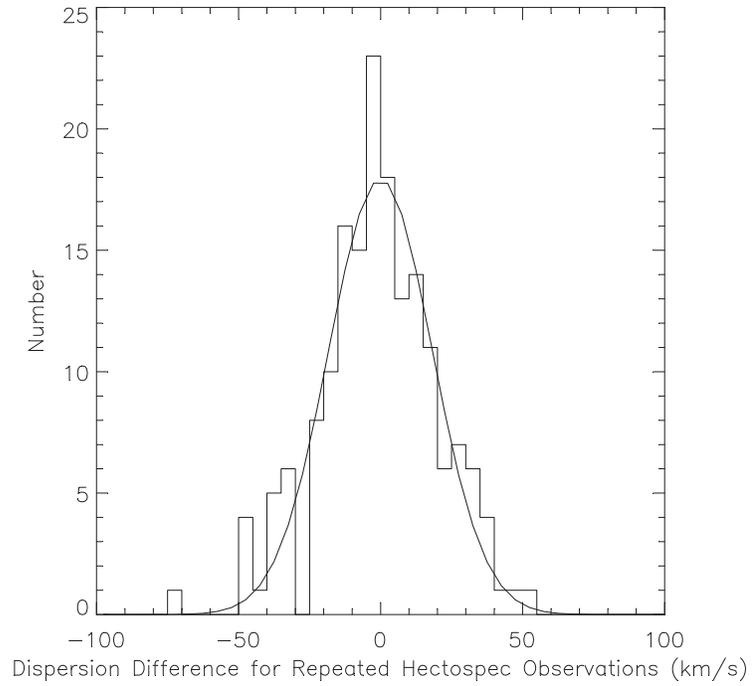}
\end{tabular}
\end{center}
\caption{\label{interr}  Hectospec internal velocity dispersion differences for repeated measurements of 171 galaxies where the error
in each measurement is $<$30 km s$^{-1}$.  The expected RMS dispersion difference calculated from the errors is 22 km s$^{-1}$ and the
measured RMS difference is 21 km s$^{-1}$.  A Gaussian of 18 km s$^{-1}$ $\sigma$ fit to the binned data is shown for reference.}
\end{figure}

We examine the internal errors in our velocity dispersions using
a sample of 171 pairs of measurements from the SHELS survey
\citep{gel10} where the error in each measurement is $<$30 km
s$^{-1}$.  Figure~\ref{interr} is a histogram of the dispersion
differences in km s$^{-1}$.  The expected RMS dispersion difference
calculated from the ULySS errors is 21.7 km s$^{-1}$, we measure
an RMS dispersion difference of 20.5 km s$^{-1}$ for the 171 pairs. We
show Gaussian of 18 km s$^{-1}$ $\sigma$ fit to the binned data for
reference in Figure~\ref{interr}.  Our repeated
measurements confirm the ULySS error estimates, and we adopt these errors
for subsequent analysis.

\section{External Errors - Comparison with SDSS/BOSS DR9}

We assess our external errors using Hectospec velocity dispersion
measurements for a sample of 984 galaxies with high SN
velocity dispersion measurements (estimated dispersion errors $<$20 km
s$^{-1}$ in both cases) in the 9th SDSS/Baryon Oscillation
Spectroscopic Survey (BOSS) data release (DR9, \citet{ahn12}); 843 of
these spectra were obtained with the SDSS fiber-fed spectrographs and
141 with the updated BOSS fiber-fed spectrographs \citep{sme13}.  The median
redshift of the combined sample is 0.147.  Galaxies with [OII]$\lambda$3727 equivalent
widths $>$5\rm\AA~were excluded from this sample to avoid contamination from
emission lines.

The Hectospec spectra are drawn from
three surveys: SHELS \citep{gel10,gel12,hwa13}, the Hectospec Cluster Survey(HeCS) \citep{rin13},
and HectoMap \citep{gel11}.  These surveys include $\sim$15,000, $\sim$22,000, and $\sim$52,000
Hectospec spectra, respectively, all with the same Hectospec configuration.
SHELS is a magnitude limited survey complete to R=20.6; here we use
the entire sample including some galaxies fainter than the R=20.6 limit.  
HeCS surveys red sequence galaxies with
r$<$21 in 58 galaxy clusters.  HectoMAP is a survey of red selected
galaxies in a 50 deg$^2$ strip to r$<$21.3.  By design, only a small fraction of 
these galaxies overlap with one of the SDSS surveys.

The SDSS-I/II fibers subtend 3$^{\prime\prime}$, the SDSS-III/BOSS fibers subtend
2$^{\prime\prime}$. To compare the Hectospec measurements directly with the DR9
velocity dispersions, we apply an aperture correction to scale
the Hectospec velocity dispersions downwards to match the larger SDSS
or BOSS apertures.  We use the aperture correction from \citet{cap06}:

$$ {{\sigma_1}\over{\sigma_2}}=\left({{r_1}\over{r_2}}\right)^{-0.066}$$

\begin{figure}
\begin{center}
\begin{tabular}{c}
\includegraphics[height=7in]{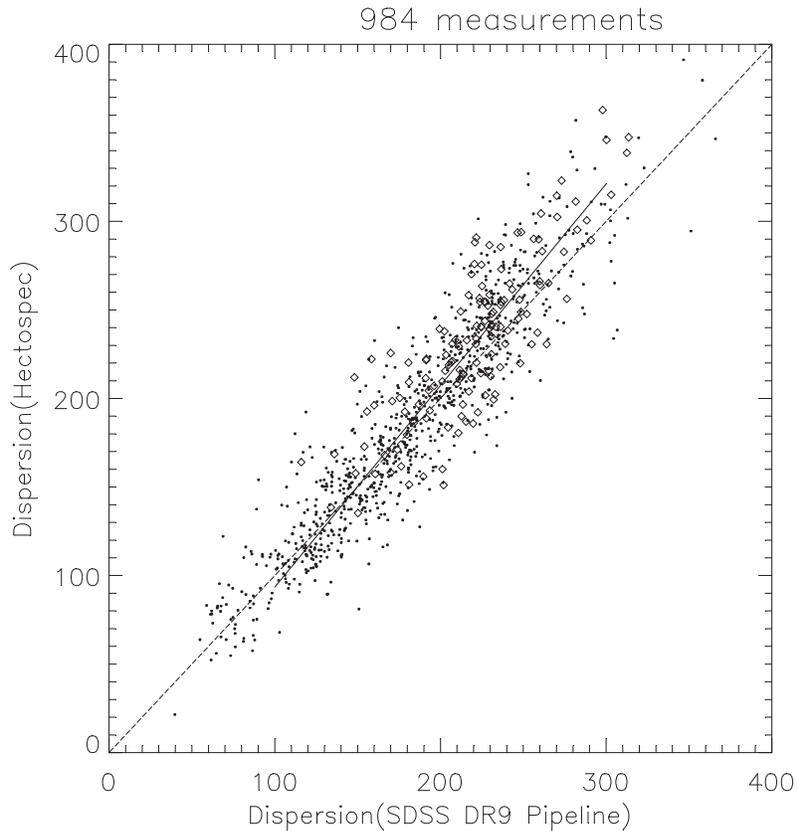}
\end{tabular}
\end{center}
\caption{\label{exterr}  A comparison of independent Hectospec and SDSS measurements of 984 galaxies with velocity dispersion errors $<$20 km s$^{-1}$.
An aperture correction has been applied to the Hectospec data (see text). The short solid line segment is a error weighted fit of a line to the plotted data
points.  This line has an intercept of -20.5 and a slope of 1.139. Measurements with the original SDSS
spectrograph and 3$^{\prime\prime}$ fibers are plotted with filled symbols (843 galaxies) and measurements with the updated BOSS spectrograph with 2$^{\prime\prime}$ fibers
are plotted with open symbols (141 galaxies).}
\end{figure}

The aperture corrections to transform to the SDSS and BOSS apertures are
thus 0.955 and 0.981, respectively.  Figure~\ref{exterr} shows
the 984 pairs of Hectospec and DR9 \citep{dr9pipe} pipeline velocity
dispersions.  We remove two galaxies from analysis with
dispersions that disagree by more than 5 $\sigma$.  Although the two
measurements of independent spectra analyzed with rather different
software show remarkable agreement, it is clear that the
Hectospec/ULySS velocity dispersions are systematically larger at
large dispersions.  The solid line in Figure~\ref{exterr} shows the
error-weighted best fit line relating the two measurements; the line
(valid only between ~100 and 300 km s$^{-1}$) has an intercept of -20.5
km s$^{-1}$ and a slope of 1.139.

We can also study the error distribution. The expected RMS dispersion
difference calculated from the ULySS and DR9 pipeline errors is 17.1
km s$^{-1}$, but we measure a larger RMS dispersion difference of 24.7
km s$^{-1}$ for the 984 pairs.  Here, we have removed the linear
relation described above.  The measured RMS dispersion difference
suggests that the errors are underestimated by $\sqrt{2}$.
However, various systematic errors like aperture centration and
aperture corrections probably contribute.  Because
our repeated measurements agree within the ULySS errors,
systematic errors may dominate the apparent excess error.

\begin{figure}
\begin{center}
\begin{tabular}{c}
\includegraphics[height=7in]{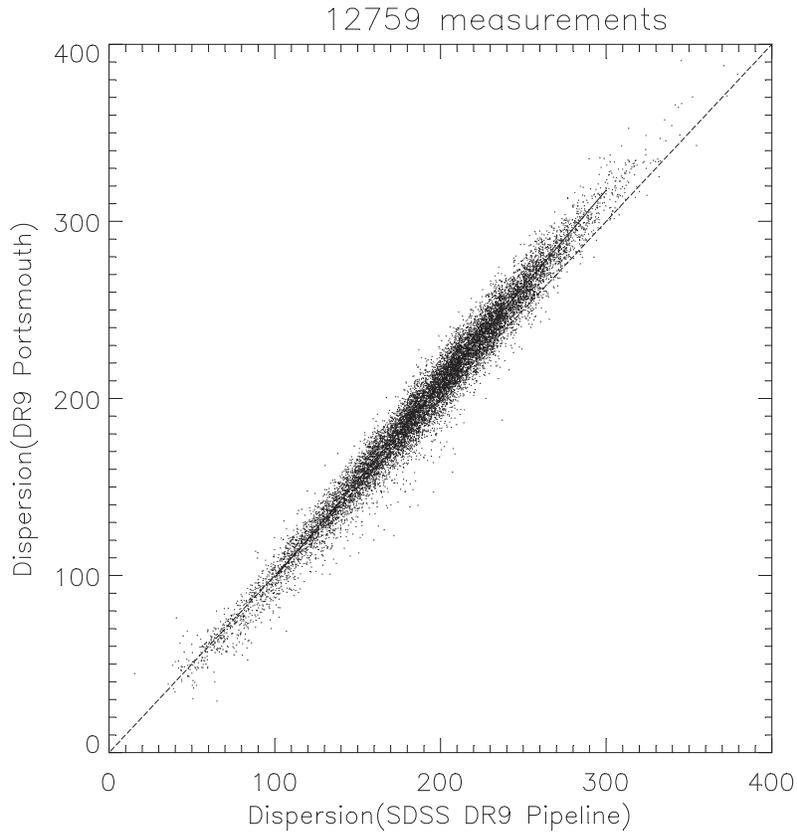}
\end{tabular}
\end{center}
\caption{\label{port}  A comparison of DR9 Portsmouth and DR9 pipeline measurements of 12759 galaxies with velocity dispersion errors $<$10 km s$^{-1}$.
The short solid line segment is a error weighted fit of a line to the plotted data points.  This line has an intercept of -10.4 and a slope of 1.093.}
\end{figure}

\begin{figure}
\begin{center}
\begin{tabular}{c}
\includegraphics[height=7in]{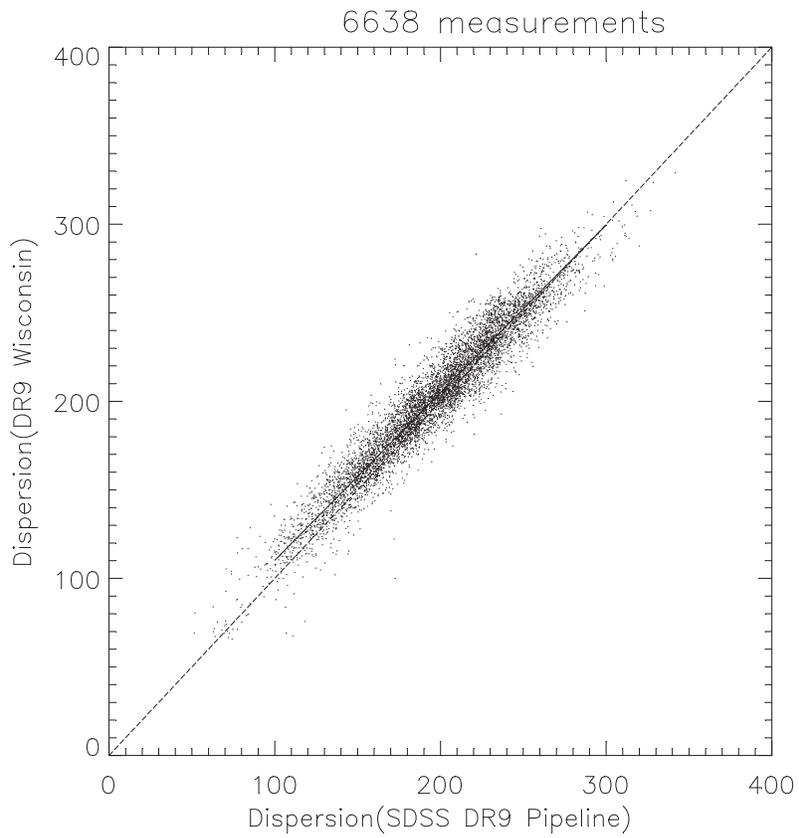}
\end{tabular}
\end{center}
\caption{\label{wisc}  A comparison of DR9 Wisconsin and DR9 pipeline measurements of 6638 galaxies with velocity dispersion errors $<$20 km s$^{-1}$.
The short solid line segment is a error weighted fit of a line to the plotted data points.  This line has an intercept of 15.6 and a slope of 0.947.}
\end{figure}

We next consider the systematic differences in velocity
dispersions.  We can explore the effect of different velocity
dispersion analysis techniques by comparing the three BOSS DR9
reductions; the pipeline reduction \citep{dr9pipe}, the Portsmouth
reduction \citep{dr9port}, and the Wisconsin reduction
\citep{dr9wisc}.  Figure~\ref{port} shows the velocity
dispersions of the subset of 12759 BOSS galaxies with velocity
dispersion errors of $<$10 km s$^{-1}$ in both analyses.  The solid
line in Figure~\ref{exterr} shows the error-weighted best fit line
relating the two measurements; the line (valid only between ~100 and
300 km s$^{-1}$) has an intercept of -10.4 km s$^{-1}$ and a slope of
1.093.  The systematic difference between the two analyses is
negligible at 100 km s$^{-1}$, rising to $\sim$6\% at 300 km s$^{-1}$.
Figure~\ref{wisc} shows the velocity dispersions of the subset of
6638 BOSS galaxies with velocity dispersion errors of $<$20 km
s$^{-1}$ in both analyses.  The solid line in Figure~\ref{exterr}
shows the error-weighted best fit line relating the two measurements;
the line (valid only between ~100 and 300 km s$^{-1}$) has an
intercept of 15.6 km s$^{-1}$ and a slope of 0.947.  The systematic
difference between the two analyses is $\sim$10\% at 100 km s$^{-1}$,
and negligible at 300 km s$^{-1}$.  \citet{dr9port} presents plots
comparing the various DR9 BOSS reductions with a less restrictive cut
on the velocity dispersion errors.

Table~\ref{dispdiff} summarizes the ratios of the velocity dispersions
from the Hectospec analysis, the Portsmouth, and the Wisconsin
reductions relative to the DR9 pipeline.  The DR9 reductions reflect
only differing analysis techniques; the underlying spectra are
identical.  It is interesting to note that the systematic deviations of
the Hectospec/ULySS velocity dispersions relative to the DR9 pipeline
are almost identical to those of the Portsmouth reductions relative to
the DR9 pipeline.  The Portsmouth analysis uses the pPXF \citep{cap04}
code that is functionally identical to ULySS and very similar stellar
population models.  In contrast, both the Wisconsin and pipeline
analyses use some form of principal component analysis.  We conclude
that the systematic differences between Hectospec and DR9 pipeline
dispersions are no larger than the differences among the three
different DR9 reductions.

\section{External Validation Using the Fundamental Plane}

Night sky features contribute differently to velocity dispersion
errors as a function of the redshift of the measured galaxy.
The MgI b triplet at low redshift is in a spectral region where the night
sky is relatively smooth and easily subtracted; at a redshift of 0.6 this feature
is well inside the ``forest'' of OH night-sky emission lines.
The direct external validation by comparison with the SDSS in section 4
does not contain a sufficient number of high redshift objects to show the presence or
absence of systematic tendencies at redshifts above z $\sim$0.2.  We perform
an indirect validation by observing the scatter and offset about the
fundamental plane as a function of redshift.

It has long been known \citep{fab76,djo87} that
early type galaxies define a narrow relation in velocity dispersion ($\sigma$),
surface brightness ($\mu$) and effective radius ($r_e$), the fundamental plane (FP).
$$ log(r_e) = a log(\sigma) + b \mu + c $$
We use this relation to test  for redshift dependant systematics in the
dispersions and their errors.  Previous studies have shown: (1) that the FP parameterization does not
vary with redshift, at least for z$\la$0.6 \citep{kel97,jor96,van96} and (2)
that the intrinsic scatter about the FP relation is stable \citep{hyd09}.

We use spectral measurements from the SHELS survey \citep{gel10} and photometric
measurements from the SDSS DR9 \citep{ahn12} to make that test.  We adopt the r
band C-model FP calibration of \cite{sau13} (a=1.041, b= 0.30 c=-7.76) which uses
cModelMag\_r and deVRad\_r, and we follow their prescription almost exactly for
correcting the input measures for aperture differences, the k correction, and
evolution.  The two differences from their analysis are: (1) that we do not correct
the measured radius of each galaxy for its ellipticity, and (2) for redshifts
from 0.5 to 0.6 we use an extrapolation of the modified k correction of \citet{chi10} used
by \cite{sau13}.  Our extrapolation assumes an elliptical galaxy spectrum.

\begin{figure}
\begin{center}
\begin{tabular}{c}
\includegraphics[height=6in]{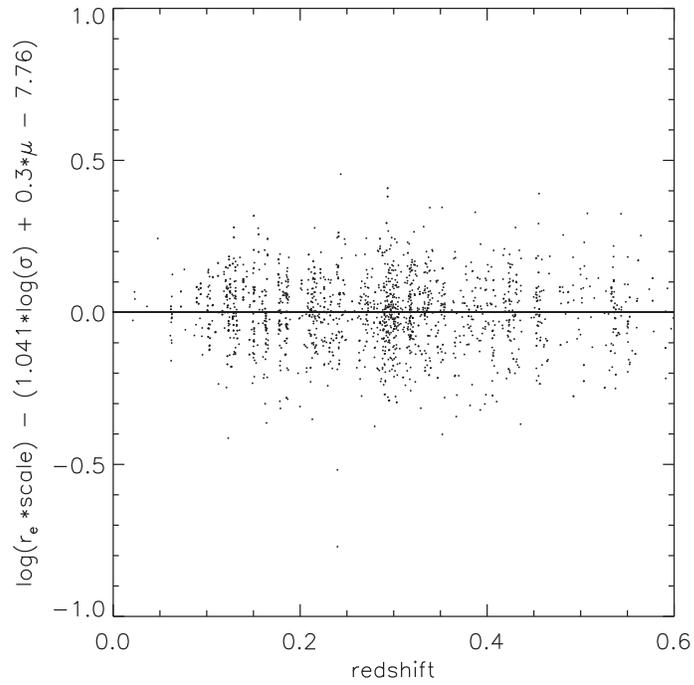}
\end{tabular}
\end{center}
\caption{\label{FP}  The scatter of 1857 galaxies from the SHELS survey \citep{gel10}
about the C-model fundamental plane relationship of \citet{sau13} as a function of redshift.
Here $r_e$ is deVRad\_r and $\mu$ is computed as $\mu = 2.5 log(2\pi) + 5 log(r_e) + cModelMag\_r$.
The sample selection is described in the text.}
\end{figure}

We choose ``red'' galaxies by requiring that deVfrac\_r $>$0.6 and OII 3727
emission $<$5 \rm\AA~and $D_{\lambda}4000$ $>$1.65.  In addition we require a
fractional error in the velocity dispersion $<$20\%.  Figure~\ref{FP}
shows the residuals in the FP as a function of redshift.  We show all 1857 objects
which pass the selection. The mean offset is 0.0011 with a
standard deviation of 0.1181.    We compute the number of objects, the mean
offset and the standard deviation in three redshift bins: (0$<$z$<$0.25; n=719;
offset=0.0077; s.d.=0.1176), (0.25$<$z$<$0.25); n=600; offset= -0.0020, s.d.=0.1195),
(0.35$<$z$<$0.60; n = 483; offset=-0.0075, s.d.=0.1211).  The offsets of the
whole sample and the redshift selected subsamples are consistent with
zero.  Thus there is no significant redshift dependent error
in the measured velocity dispersions.

We externally validate the errors in the velocity dispersion measurements by
computing the intrinsic scatter in the FP with two different cuts in fractional
velocity dispersion error.  Because we compute the intrinsic width of the FP
by subtracting (in quadrature) the error expected by propagating the individual
measurement errors from the measured scatter, under(over) estimates of the
measurement errors will produce a computed intrinsic width that differs for
samples with different error cuts.

To eliminate any redshift dependent effects we narrow our analysis to the central
redshift bin: 0.25$<$z$<$0.35.  We compute the FP intrinsic width from two samples:
(1) galaxies with a maximum fractional dispersion error of 0.20, and (2)
galaxies with a fractional dispersion error $<$0.10.  For the 20\%\ error
sample we compute an intrinsic width of 0.0999 from 600 galaxies; for the 10\%\
error sample we compute an intrinsic width of 0.1008 from 182 galaxies.  The
difference in the computed widths is small. If the difference resulted entirely from a misestimation
of the velocity dispersion errors, the errors would have to be overestimated by
5\%, in basic agreement with the internal error analysis in section 3.

\section{Planning Hectospec Dispersion Measurements}

Here we compute the velocity dispersion errors expected as a function of
magnitude for an exposure time of 3600 s.  We have insufficient experience with
significantly longer exposures to determine the errors as a function
of exposure time.

\begin{figure}
\begin{center}
\begin{tabular}{c}
\includegraphics[height=7in]{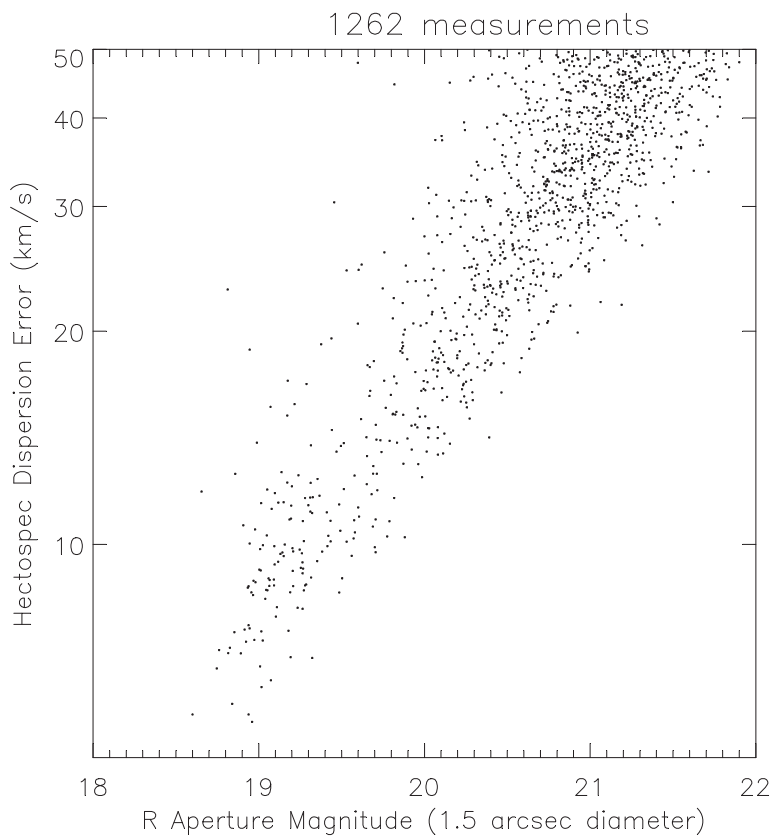}
\end{tabular}
\end{center}
\caption{\label{errmag} Hectospec velocity dispersion errors for a 3600 s observation as a function of R magnitude within a 1.5$^{\prime\prime}$
Hectospec fiber aperture.  The points reflect observations of galaxies with velocity dispersions between 100 and 300 km s$^{-1}$ during a wide
range of conditions including seeing and transparency.  Although we expect the errors to depend on the galaxy dispersion, this dependence is
obscured by the seeing and transparency variations that we cannot accurately remove.  The lower envelope reflects observations during the
best conditions.}
\end{figure}

We examine a sample of 1262 SHELS galaxies observed for
3600 s with velocity dispersion errors $<$50 km s$^{-1}$.  We expect the errors
to correlate best with 1.5$^{\prime\prime}$ aperture magnitudes corresponding to
the Hectospec fiber diameter.  Figure~\ref{errmag} plots the Hectospec velocity
dispersion errors as a function of R magnitude in a 1.5$^{\prime\prime}$ aperture
\citep{wit02,wit06}.
The correlation reflects a range of observing conditions including seeing, cloud
cover, and moon illumination.  The lower envelope of the distribution corresponds
to observations during dark conditions with good seeing and clear skies.  Under
these conditions, with a 3600 s observation we expect a 20 km s$^{-1}$ velocity
dispersion error for
a galaxy with an R aperture magnitude of 20.5, and a 30 km s$^{-1}$ velocity
dispersion error for a galaxy with an R aperture magnitude of 21.

\begin{figure}
\begin{center}
\begin{tabular}{c}
\includegraphics[height=7in]{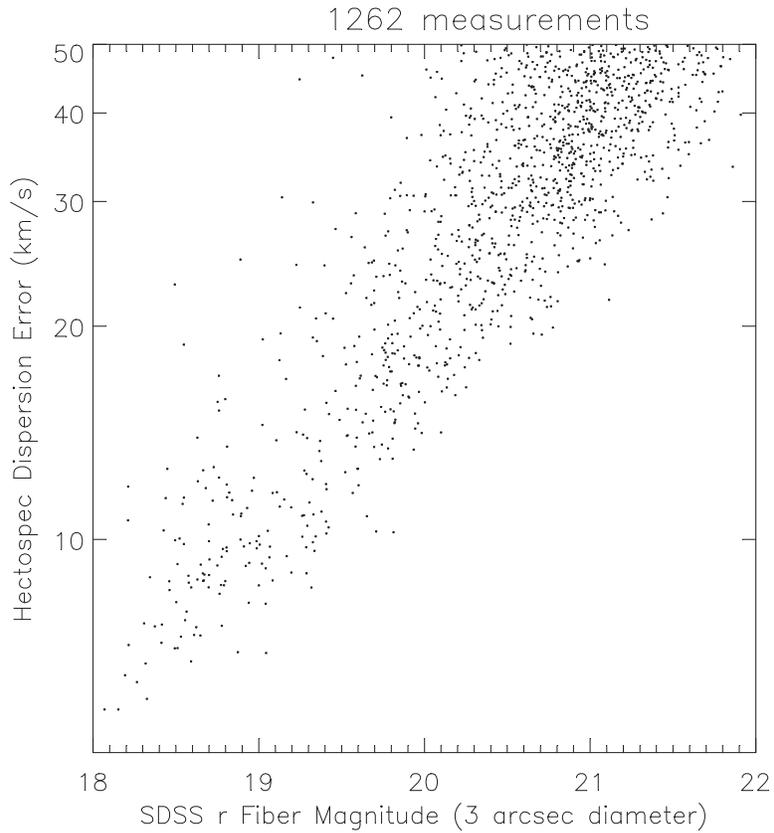}
\end{tabular}
\end{center}
\caption{\label{errmagsdss} Hectospec velocity dispersion errors for a 3600 s observation as a function of r magnitude within a 3$^{\prime\prime}$
aperture (SDSS fibermag).  The points reflect observations during a wide range of conditions including seeing and transparency.  The lower
envelope reflects observations during the best conditions.}
\end{figure}

For most observers, the best easily available proxy for Hectospec fiber magnitudes
is SDSS fiber magnitudes measured in a 3$^{\prime\prime}$ aperture.  These
magnitudes do not correlate as well with the Hectospec velocity dispersion errors as the
1.5$^{\prime\prime}$ aperture magnitudes (Figure~\ref{errmag}), but they are still useful
(Figure~\ref{errmagsdss}).  Under the best conditions, we expect a 20
km s$^{-1}$ velocity dispersion error for a galaxy with an r SDSS fiber magnitude
of 20.4, and a 30 km s$^{-1}$ velocity dispersion error for a galaxy with an
r fiber magnitude of 21.  The differences in filter bandpass and aperture
roughly cancel.

\section{Conclusions}

The main goal of our investigation is to enable studies of fundamental galaxy properties
and their evolution using Hectospec data.  Careful comparisons of velocity dispersion measurements
made with independent instruments are also of more general interest to establish the
accuracy of our large data sets in an era of ``precision cosmology".

We describe the use of publicly available software, ULySS \citep{kol09}, to obtain
velocity dispersions for Hectospec galaxy spectra.  We compare our measurements
to those from the SDSS DR9 pipeline for 984 galaxies in common with velocity dispersion
errors of $<$20 km s$^{-1}$.  The systematic differences in the
two measurements are $<$7\% for galaxies with dispersions between 100 and 300 km s$^{-1}$.
These differences are comparable to the systematic differences among
the three velocity dispersion reductions for the DR9 BOSS data.

By analyzing the scatter about the fundamental plane we show that there are no significant
systematics in our velocity dispersion measures as a function of redshift, for z$\la$0.6.
Additionally we confirm that our estimation of the measurement errors is correct, within
narrow tolerances.

In one hour in good conditions,
we can expect 20 km s$^{-1}$ velocity dispersion errors for a galaxy with an r SDSS fiber magnitude
of 20.4, and 30 km s$^{-1}$ velocity dispersion errors for a galaxy with an
r fiber magnitude of 21.

{\it Facility:} \facility{MMT (Hectospec)}

\section{ACKNOWLEDGMENTS}

We thank Marijn Franx, Nelson Caldwell and Daniel Eisenstein for helpful comments.

Observations reported here were obtained at the MMT Observatory, a joint facility of the Smithsonian Institution
and the University of Arizona.
We thank the Hectospec engineering team including Robert Fata, Tom Gauron, Marc Lacasse, Mark Mueller, and Joe Zajac,
and the instrument specialists Perry Berlind and Michael Calkins.  We are grateful for the contributions of the members
of the CfA's Telescope Data Center including Warren Brown, Anne Matthews, John Roll, Susan Tokarz and Sean Moran.
The entire staff of the MMT Observatory under the direction of G. Grant Williams has provided outstanding support for
Hectospec operations.  Nelson Caldwell has ably scheduled Hectospec's queue operations.

Funding for the SDSS and SDSS-II has been provided by the Alfred P. Sloan Foundation, the Participating Institutions,
the National Science Foundation, the U.S. Department of Energy, the National Aeronautics and Space Administration,
the Japanese Monbukagakusho, the Max Planck Society, and the Higher Education Funding Council for England. The SDSS
Web Site is http://www.sdss.org/.  The SDSS is managed by the Astrophysical Research Consortium for the Participating Institutions.
Funding for SDSS-III has been provided by the Alfred P. Sloan Foundation, the Participating Institutions, the National Science
Foundation, and the U.S. Department of Energy Office of Science. The SDSS-III web site is http://www.sdss3.org/.


\clearpage

\clearpage

\clearpage

\clearpage

\clearpage

\clearpage

\clearpage

\clearpage

\clearpage

\begin{deluxetable}{rc}
\tabletypesize{\footnotesize}
\tablecaption{Hectospec Line Spread Function (km s$^{-1}$) \label{lsf}}
\tablewidth{0pt}
\tablehead{
\colhead{Polynomial Term} & \colhead{Coefficient}
}
\startdata
constant &  4.64400 $\times$ 10$^{2}$\\
linear   & -1.15515 $\times$ 10$^{-1}$\\
quadratic&  1.16604 $\times$ 10$^{-5}$\\
cubic    & -3.99359 $\times$ 10$^{-10}$\\
\enddata
\end{deluxetable}

\begin{deluxetable}{cccc}
\tabletypesize{\footnotesize}
\tablecaption{Dispersion Ratios\label{dispdiff}}
\tablewidth{0pt}
\tablehead{
\colhead{DR9 Pipeline Dispersion(km s$^{-1}$)} & \colhead{Hectospec Ratio} & \colhead{Portsmouth Ratio}
& \colhead{Wisconsin Ratio}
}
\startdata
100& 0.93 & 0.99 & 1.10\\
150& 1.00 & 1.02 & 1.05\\
200& 1.04 & 1.04 & 1.02\\
250& 1.06 & 1.05 & 1.01\\
300& 1.07 & 1.06 & 1.00\\
\enddata
\end{deluxetable}

\end{document}